# Beat Pressure and Comparing it with Ascending Aorta Pressure in Normal and Abnormal Conditions

O. Ghasmelizadeh[1], M.R. Mirzaee[1], B. Firoozabadi[1], B. Sajadi[1], A. Zolfonoon[2]

[1] Department of Mechanical Engineering, Sharif University of Technology, Tehran, Iran
[2] Centers of Workshops, Sharif University of Technology, Tehran, Iran

*Abstract* — **Lumped method (Electrical analogy) is a quick and easy way to model human cardiovascular system. In this paper Lumped method is used for simulating a complete model. It describes a 36-vessel model and cardiac system of human body with details that could show hydrodynamic parameters of cardiovascular system. Also this paper includes modeling of pulmonary, atrium, left and right ventricles with their equivalent circuits. Exact modeling of right and left ventricles pressure with division of ascending aorta into 27 segments increases the accuracy of our simulation. In this paper we show that a calculated pressure for aorta from our complex circuit is near to measured pressure by using advanced medical instruments. Also it is shown that pressure graph from brachial is so near to aortic pressure because of this its pressure signal is usable instead of aortic pressure. Furthermore, obstruction in ascending aorta, brachial and its effects has been showed in different figures.**

## I. Introduction

Human faults of cardiovascular system, which is the main reason for diseases, are the main causes of health problems in all societies. To analyze cardiovascular system and effects of diseases on it different ways are usable such as lumped model, one or multi-dimensional modeling and experimental methods. In this context lumped method were used with the goal of providing better understanding and simulation of the blood flow in the human cardiovascular system which will lead to accurate answers as a result of exact modeling. In this simulation a three-parametric model of heart muscles behaviors was introduced and applied[1]. It accurately showed pressure development during a time period of heart action. Finite element methods were been commonly utilized to simulate and model the left function. The simulation of steady state and transient phenomena with electronic-aided circuit has also been considered but the arteries have been included partly in their model[2]. The first computer models describing the arterial system such as ascending aorta and carotids were introduced a multi-branched model of the arterial tree in a usable form for digital computers. By this method different physiological conditions became considerable. Later more detailed models were applied to reach more accurate results[3]. To analyze the human cardiovascular system mathematically, more simplified model should be considered to decrease the difficulty of investigating. A pulsatile-flow model of the left and right ventricles (as suppliers) and 2-segment aorta were constructed and the changes in flow behavior investigated. Later lumped (electrical analogy) model was developed to analyze cardiovascular systems easily with suitable accuracy. An electrical model which focused on the vessel properties was made by Young and his team[4]. One-dimensional axisymmetric Navier-Stokes equations for time dependent blood flow in a rigid vessel had been used to derive lumped models relating flow and pressure[5]. The effect of drugs on the blood properties and its circulation was studied later in a 19-compartment model[6]. A complicated non-linear computer model for pressure changes and flow propagation in the human arterial system was drived[7]. The model had 55 arterial compartments and was based on one-dimensional flow equations to simulate effects of hydrodynamic parameters on blood flow. A Computational model was developed to provide boundary conditions for simulations of the effects of endoleak on the AAA wall stress[8]. Later a simple model of lumped method was presented for human body[9].

This paper describes modeling of the whole human cardiovascular system using a extensive equivalent electronic circuit. In this method we have taken a quite different way to model the system and we had intention to develop our electrical models by addition more segments to main arteries which more accurate results would be one of its effects. The model consists of about 120 RLC segments representing the arterial and cardiac systems. Compared to previous studies, both arterial and cardiac systems of this model are more detailed, especially ascending aorta and upper arteries such as carotid, brachial and Brachiocephalic. So more exact results for ascending aorta and other named vessels properties could be reachable. Logically the method of increasing the number of segments for each artery will increase the accuracy of answers greatly. Also investigating cardiovascular system faults and effects of diseases would cause precise results. In addition using more compartments to model each artery creates this possibility to simulate variable radius of each artery quite exactly. Therefore because of considering variable radius, abnormalities such as obstructions in different arteries can be probed quite real. Another advantage of this way of modeling is that we can consider different elasticity modulus (E) for each segment, so investigating effects of this variation in human cardiovascular system will be possible.

Also, it is obvious that aortic and brachial pressure are so similar to each other because of this it is so easiest to take brachial pressure instead of aortic pressure. In this article by cardiovascular modeling with equivalent circuit which will be explained below we reached to this similarity, and this is what we can use to understand good or bad working of human cardiovascular system just by taking the beating pressure of a person. Increasing more elements to our circuit was so productive to reach these similarities. Pressure graphs which are shown in verification part would confirm the above speeches.

Here is a 36-vessel body tree which is utilized in this research to model human cardiovascular system vessels.

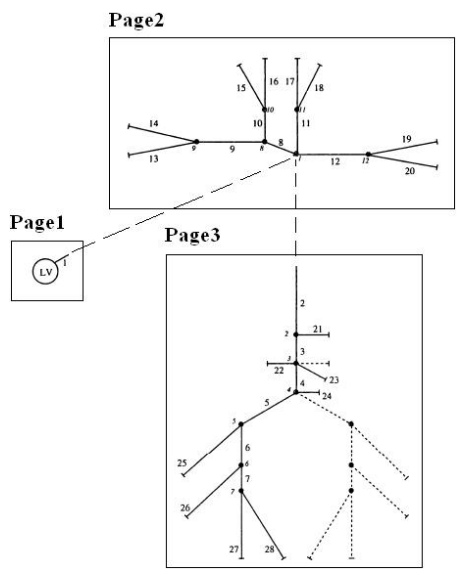

Fig. 1 – 36-Vessels Body Tree

## II. MODELING PRINCIPLES

To model human cardiovascular system we chose different equivalent electrical elements to express different mechanical properties of vessels, blood and heart.

In our modeling process atriums, ventricles, every blood vessel, set of all capillaries, arterioles and veins have been presented by some compartments consisting of a resistor, an inducer and a capacitor.

The number of compartments would be chosen by the purpose and the required accuracy. So more compartments would be used for main arteries.

Voltage, current, charge, resistance and capacitance, inductor in the electronic circuit are respectively equivalent to blood pressure, blood flow, volume, resistance, compliance and flow inertia in the cardiovascular system. Ground potential (reference for voltage measurements) is assumed to be zero as usual. The relation between mechanical properties of cardiovascular system and their equivalent electrical elements are as follow:

0.01ml/Pa = 1 µF (compliance - capacitance)
1 $Pa.s^2$/ml = 1 µH (inertia - inductor)
1 Pa.s/ml = 1 kΩ (resistance - resistance)
1mmHg = 1 volt (pressure - voltage)
133416 ml = 1A (volume - charge)

Blood vessel resistance (R), depending on blood viscosity and vessel diameter, is simulated by resistors. This simulation has considered because blood viscosity will cause resistance against Blood flow crossing.

The blood inertia (L) is simulated by inductors. Reason of this consideration is variability of flow acceleration in pulsatile blood flow, so an inductor can model inertia of blood flow very clearly.

The vessel compliance (C) is considered using capacitors. For the reason of this simulation, it should be noted that by passing blood thorough vessels, the vessels would be expanded or contracted, so they can keep blood or release it, and this is exactly like what a capacitor does. By these statements each vessel is modeled by some compartments, which includes one resistance, one capacitor, and one inductor. Quantities of Compartment's elements are easily achievable by using equations from physiological book[10]. Computed values of circuit Elements are shown in table 1.

Where n is number of each artery segments.

Also, artery wall thickness (h) is obtained from fig2. (Physiological text[11])

Atriums are simulated as part of the venous system without any contraction. Atriums and ventricles can be modeled like vessels as a RLC segment. These two important parts of heart have two courses of action, one is resting position (diastole) which muscles would take their maximum volume and the other is acting position (systole) which muscles reach to their minimum size.

Also we should note that heart with its moving muscles is the power supplier of blood circulation in whole body vessels.

To have a very exact modeling of blood and vessels behaviors, exact modeling of power suppliers is an important factor. So in our simulation left and right ventricles are modeled quite exactly the same as biological graph sources.

Table 1 – Calculated Values for Elements of Circuit from artery parameters

| Number | Vessel Name | A (cm^2) | l (cm) | n | E (Mpa) | R (KΩ) | C (µF) | L (µH) |
|---|---|---|---|---|---|---|---|---|
| 1 | Ascending Aorta | 6.60507 | 0.204 | 27 | 0.4 | 0.000411 | 0.044883 | 0.007286 |
| 2 | Thoracic Aorta | 3.59701 | 18.5 | 1 | 0.4 | 0.125776 | 2.15337 | 1.215072 |
| 3 | Abdominal Aorta | 2.378 | 4.3 | 1 | 0.4 | 0.066889 | 0.303284 | 0.427197 |
| 4 | Abdominal Aorta | 1.021 | 9.6 | 1 | 0.4 | 0.810079 | 0.261925 | 2.221352 |
| 5 | Common iliac | 0.849 | 19.2 | 1 | 0.4 | 2.343114 | 0.418125 | 5.342756 |
| 6 | Femoral Artery | 0.181 | 43.2 | 1 | 0.8 | 115.9936 | 0.073314 | 56.38674 |
| 7 | Anterior Tibial Artery | 0.053 | 1.5 | 1 | 1.6 | 46.97287 | 0.000277 | 6.686321 |
| 8 | Brachiocephalic | 1.20798 | 2.4 | 1 | 0.4 | 0.144677 | 0.079312 | 0.469379 |
| 9 | R Brachial | 0.50298 | 5.125 | 8 | 0.4 | 1.78196 | 0.060436 | 2.40721 |
| 10 | R Common Carotid | 0.50298 | 16.8 | 1 | 0.4 | 5.841352 | 0.198113 | 7.890951 |
| 11 | L Common Carotid | 0.50298 | 11 | 1 | 0.4 | 3.824695 | 0.129717 | 5.166694 |
| 12 | L Brachial | 0.55399 | 44.4 | 1 | 0.4 | 12.72576 | 0.586882 | 18.9344 |
| 13 | R Radial | 0.08 | 4.64 | 5 | 0.8 | 63.76945 | 0.002583 | 13.70198 |
| 14 | R Ulnar | 0.13901 | 4.58 | 5 | 0.8 | 20.8499 | 0.005343 | 7.783999 |
| 15 | R External Carotid | 0.196 | 11.3 | 1 | 0.8 | 25.87461 | 0.021169 | 13.62054 |
| 16 | R Internal Carotid | 0.283 | 17.2 | 1 | 0.8 | 18.89136 | 0.050727 | 14.35866 |
| 17 | L Internal Carotid | 0.283 | 17.2 | 1 | 0.8 | 18.89136 | 0.050727 | 14.35866 |
| 18 | L External Carotid | 0.196 | 11.3 | 1 | 0.8 | 25.87461 | 0.021169 | 13.62054 |
| 19 | L Radial | 0.13901 | 4.64 | 5 | 0.8 | 21.12304 | 0.005413 | 7.885973 |
| 20 | L Ulnar | 0.08 | 4.58 | 5 | 0.8 | 62.94484 | 0.00255 | 13.5248 |
| 21 | Coeliac | 0.478 | 1 | 1 | 0.4 | 0.384992 | 0.010925 | 0.494247 |
| 22 | Renal | 0.212 | 2.7 | 1 | 0.4 | 5.284447 | 0.01138 | 3.008844 |
| 23 | Sup Mesenteric | 0.581 | 5.4 | 1 | 0.4 | 1.407178 | 0.07666 | 2.195783 |
| 24 | Inf Mesenteric | 0.08 | 4.5 | 1 | 0.4 | 61.85005 | 0.00501 | 13.28906 |
| 25 | Profundis | 0.166 | 12.1 | 1 | 1.6 | 38.62573 | 0.00921 | 17.22063 |
| 26 | Post Tibial | 0.102 | 30.6 | 1 | 1.6 | 258.7192 | 0.012262 | 70.875 |
| 27 | Ant Tibial | 0.031 | 29.5 | 1 | 1.6 | 2700.264 | 0.002839 | 224.8185 |
| 28 | Proneal | 0.053 | 31.3 | 1 | 1.6 | 980.1671 | 0.005771 | 139.5212 |

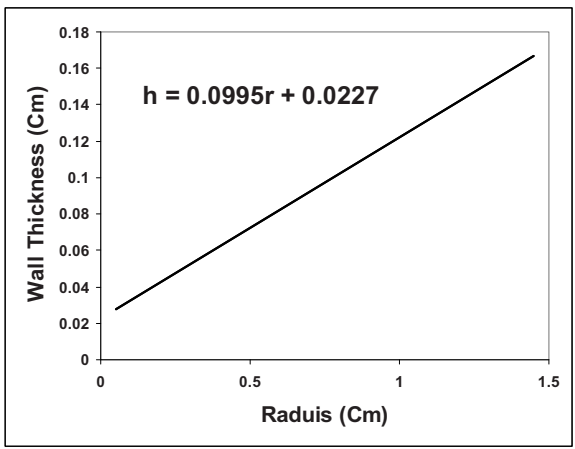

Fig. 2 – Wall Thickness of Vessels versus Radius

Different heart shutters are modeled by using appropriate diodes, because shutters like diodes cross the flow in one direction. Considering this fact is important because in some parts of cardiovascular system, inverse current movement will cause to great danger to health. So to reach a good model, choosing appropriate diodes would be counted as an inseparable part. Type of diodes is visible in circuit at Fig3.

Also, bifurcations are important cases to have accurate modeling. For simulating these parts of cardiovascular system a special method has been used[10]. In brief, it can be said that in bifurcations properties of jointed vessels will combine together in a complicated way to show the blood current division effects.

### III. OBSTRUCTION MODELING

Obstructions in important places would be modeled. In this paper constriction in ascending aorta and right brachial would be had in mind.

To have an exact obstruction model, we should note that this phenomenon doesn't occur in one point very sharply but it happens smoothly, it means the radius will decrease slowly and in one point it reaches to its minimum quantity, after that it begins to increase again to its first quantity (vessel radius without obstruction). So for each artery with respect to its length and its importance, numbers of segments would be chosen. These segments will show the changes of radius in the region of obstruction.

As said in physiological text[11], ascending aorta is 5.5 centimeters and in our circuit it has been modeled with 27 compartments. By above considerations, length of each compartment becomes 2.037 millimeters. So to model obstruction region we considered 1.185 centimeters length of ascending aorta that includes 5 compartments with numbers 1, 2, 3, 4, and 5. From the same physiological text[11] it is obvious that right brachial length is 41 centimeters and in equivalent circuit it has been modeled with 8 compartments which each segments length become 5.125 centimeters. To model obstruction for this artery 1 segment would be chosen with 60% constriction which it is significant with number 6 in equivalent circuit. One segment is chosen because obstruction would not happen with much decreasement in radius along brachial vessel.

In order to calculate obstruction elements values it should note that capacitor quantity relates to wall thickness and elasticity modulus. Values of these parameters would vary because of obstruction. Wall thickness (h) increases with respect to percent of obstruction and also, elasticity modulus (E) variation would be computable with considering equation 1.

$$a = \sqrt{\frac{Eh}{2r\rho}} \qquad (1)$$

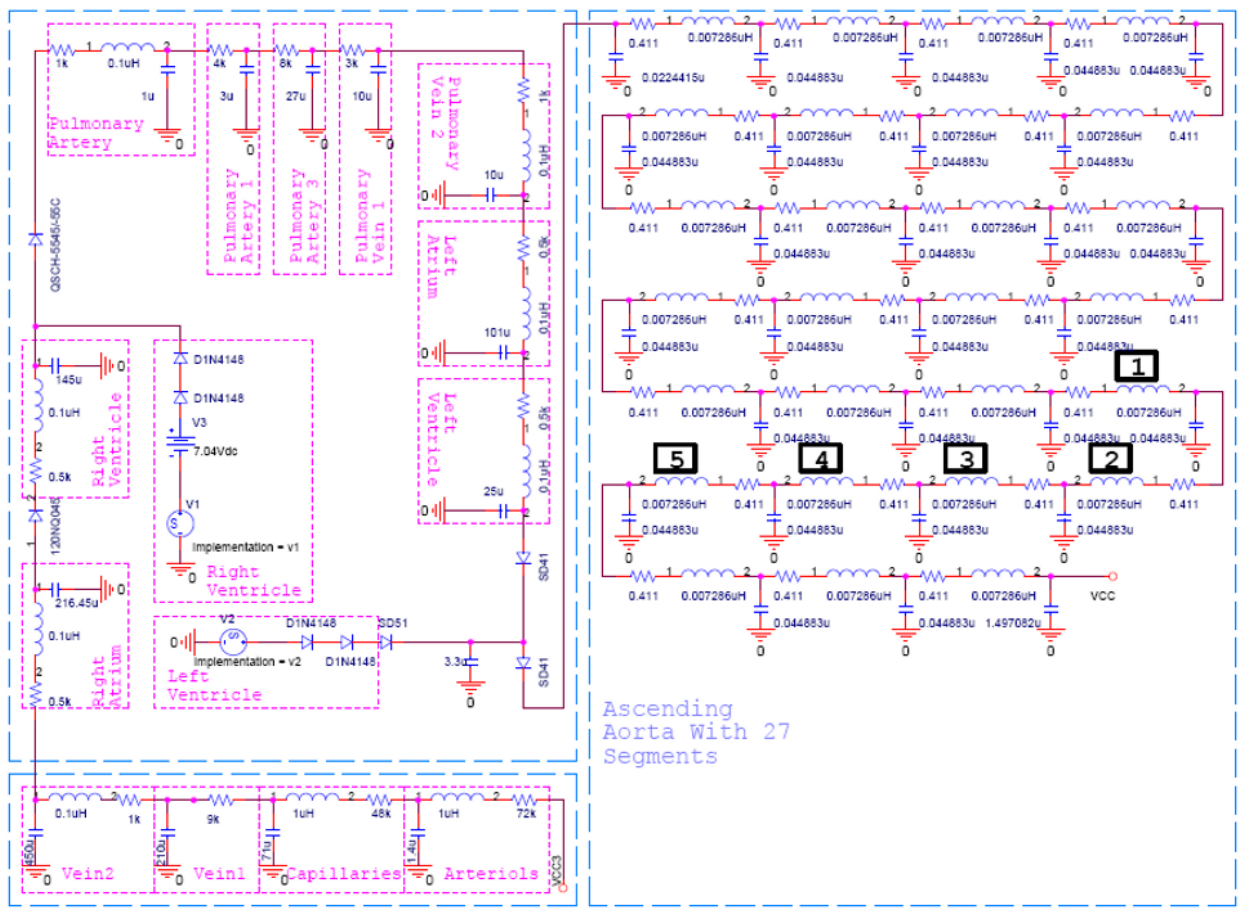

Fig. 3 – Circuit of Heart and Ascending Aorta

Where (a) is wave propagation velocity in vessel.

Note that (a) would remain constant because there is no change in vessel material. The new computed values for radius (r), thickness (h), and elasticity module (E) are given in table 2.

By referring to table 2, obstruction elements such as R, L, and C would be as follow in table 3.

Also equivalent figures and circuits for obstruction in both named arteries are visible in figure 6.

Table 2 – Calculated Values for Vessel Properties in Obstruction

| Obstruction Percentage | | $r_2$ (Cm) | $E_2$ (Mpa) | $h_2$ (Cm) |
|---|---|---|---|---|
| Ascending Aorta | 90 | 0.144999 | 0.00444 | 1.468 |
| | 60 | 0.579995 | 0.02525 | 1.033 |
| | 30 | 1.014991 | 0.07632 | 0.598 |
| Brachial | 60 | 0.160052 | 0.03368 | 0.304 |

Table 3 – Calculated Values for Obstruction Elements

| Obstruction Percentage | | R (Ω) | C (µF) | L (µH) |
|---|---|---|---|---|
| Ascending Aorta | 90 | 4107.25 | 0.72861 | 0.003815 |
| | 60 | 16.0439 | 0.04554 | 0.014587 |
| | 30 | 1.71064 | 0.01487 | 0.033438 |
| Brachial | 60 | 69607.8 | 0.00484 | 15.04506 |

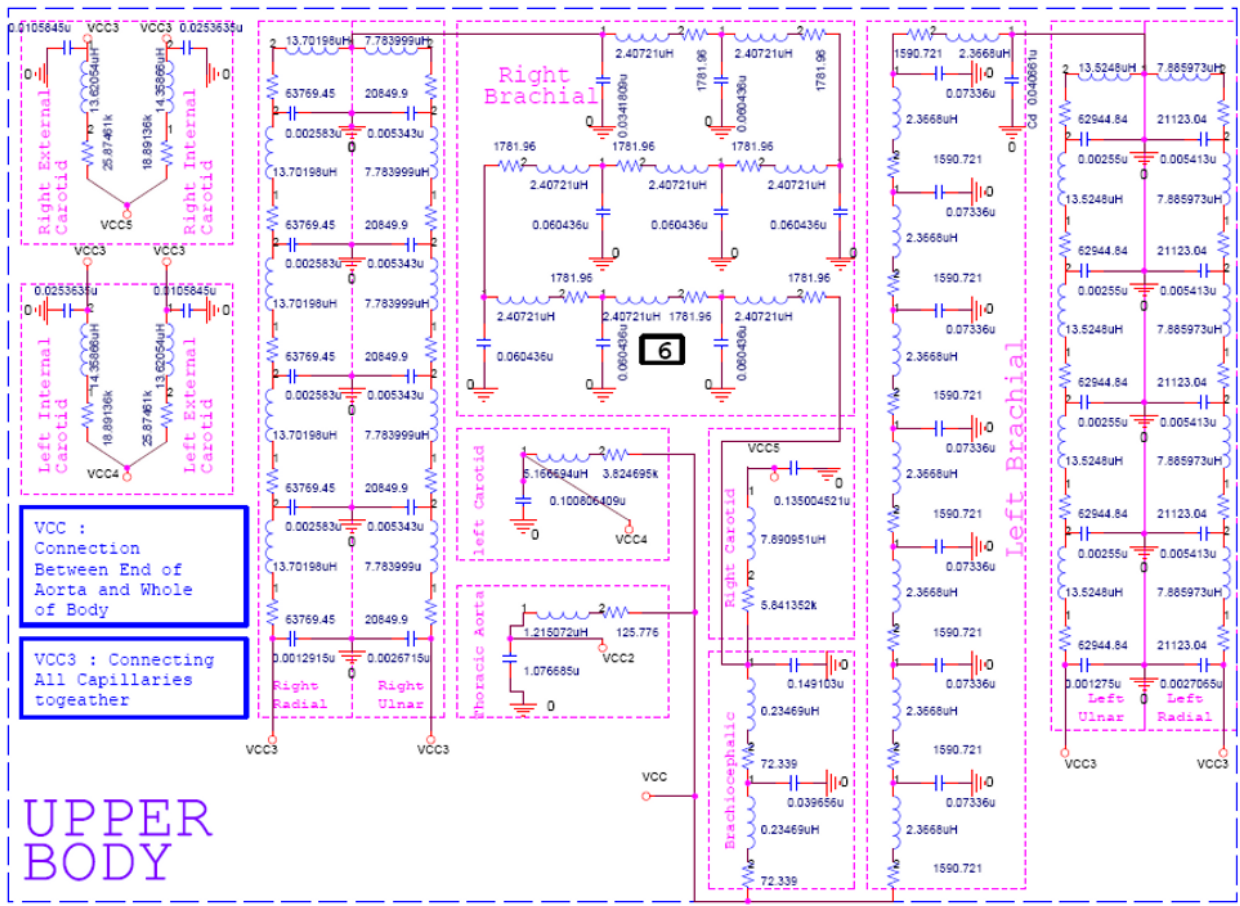

Fig. 4 – Circuit of Upper Body including Hands and Carotids

## IV. CIRCUIT DESCRIPTION

Human Cardiovascular system equivalent circuit which its elements quantities have been earned from modeling principles is shown in Figs 3, 4, and 5. This circuit includes three main circuits that each one shows one part of body would be as follow.

- Ascending aorta and heart (ventricles, atriums, pulmonary and shutters).

The complex circuit is shown in Fig 3.

Ascending aorta would be subdivided into 27 segments which elements quantities are shown in table 1.

The right atrium and ventricle are modeled by two capacitors 216.45 μF and 150 μF. Also the left atrium and ventricle are represented by two capacitors 101 μF and 25 μF [10].

As said before we should use appropriate diodes to model shutters. Diodes used for tricuspid, pulmonary, mitral and aortic shutters are in respect 120NQ045, QSCH5545/-55C, SD41 and SD41.

- upper part of body (hands and carotids)
- downer part of body (thoracic aorta and feet)

Complex circuit of upper and downer body is shown in Fig4 and Fig5.

By seeing the number of segments in these two parts the accuracy of our modeling would be more proved.

In human body Cardiac current output and aortic pressure should be in turn 100 (ml/s) and 120-68 (mmHg) [12]. These were exactly calculated in our circuit, so our modeling correctness is verifying undoubtedly.

Also, equivalent circuits for obstructions in ascending aorta and right brachial are shown in figure 6. A modeled shape of obstruction for each of these arteries is shown below the circuits.

It should be noted, because each segment in obstruction region has its own radius, its elements quantities would be different and we should use formulas 1, 2 and 3 separately to compute their amounts.

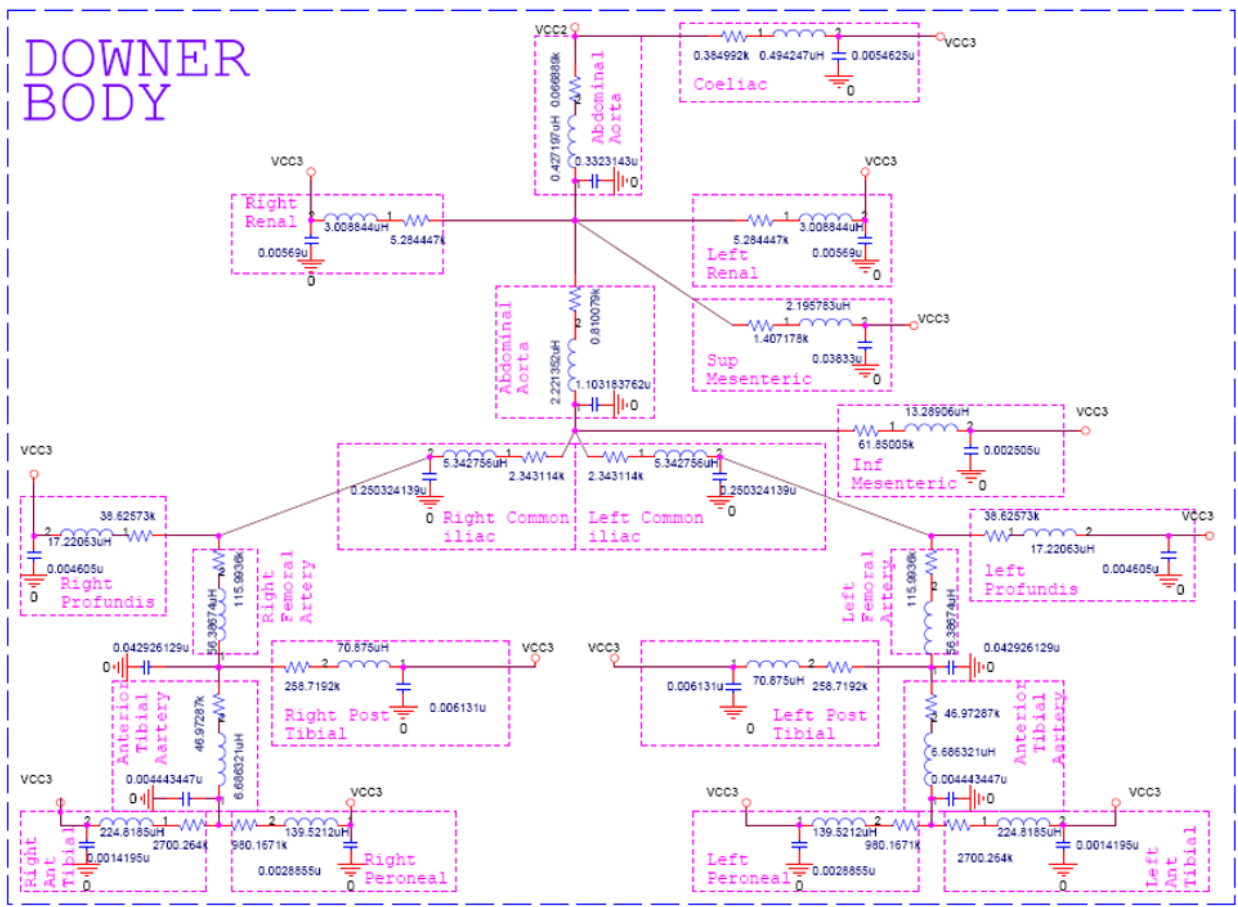

Fig. 5 – Circuit of Lower Body including Descending Arteries

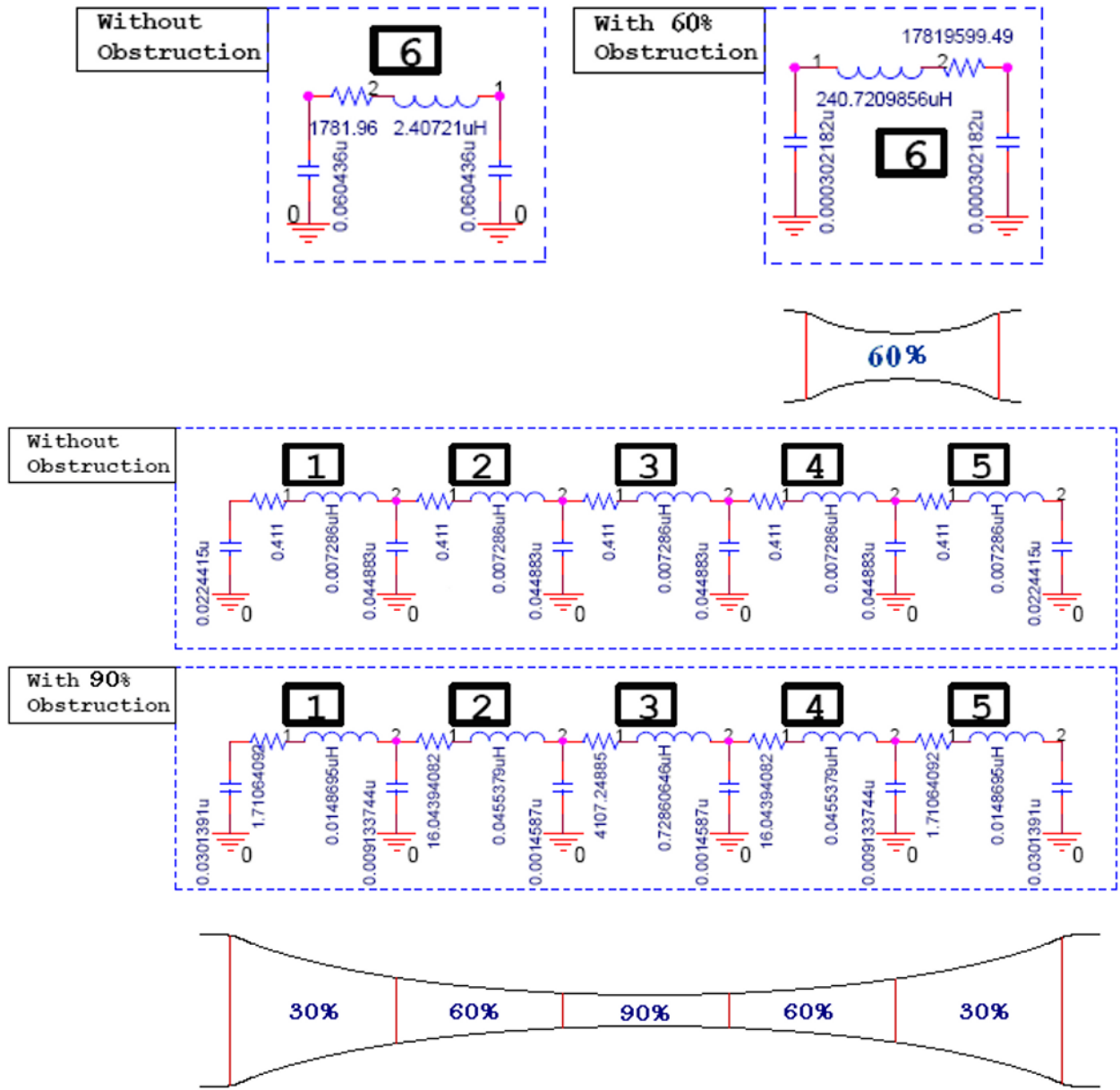

Fig. 6 – Circuit and Shape of Obstruction in Brachial and Ascending Aorta

One of the most important points in making these circuits is that arteries should end to arterioles and capillaries. Also, capillaries should be connected to veins that are the last part of blood circulation and are connected to heart at last. In our circuit marker "Vcc3" will play capillaries and arterioles role.

So by completing circuit using capillaries, arterials and veins the generated current of left ventricle distributes blood through ascending aorta to all arteries. So blood continues its path toward the body arteries and then arterioles and capillaries. Finally by using veins blood comes back to heart. In heart, right atrium receives whole blood and sends it to right ventricle. This part by contracting will send the current to pulmonary. After purifying, left atrium receives blood and sends it through mitral valve to left ventricle which is the main supplier and pumps blood into the whole body again.

Model is capable of showing blood and vessels properties. For example the pressure (voltage) and volume (charge) graphs can be obtained from the different points of the circuit easily.

It should be noted that by using more than 120 segments for whole arteries there is no leakage of current in the system and the arteries pressure graph are quite acceptable comparing to the input voltage.

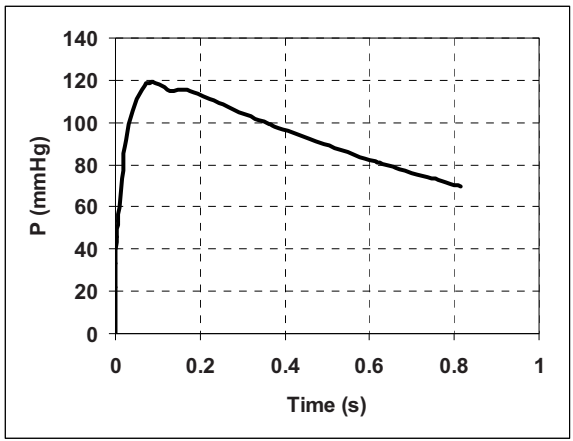

Fig. 7 – Calculated Pressure for Aorta

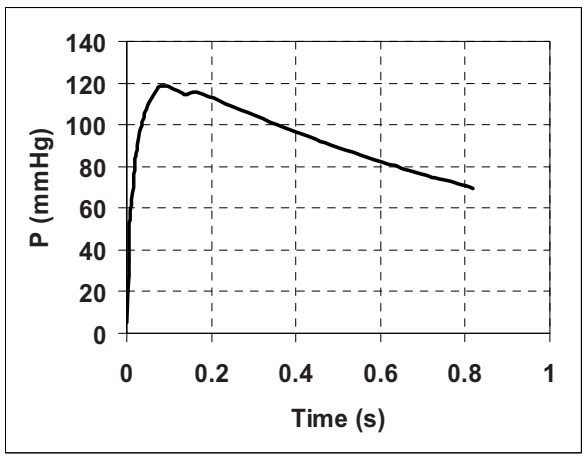

Fig. 8 – Calculated Pressure for Brachial

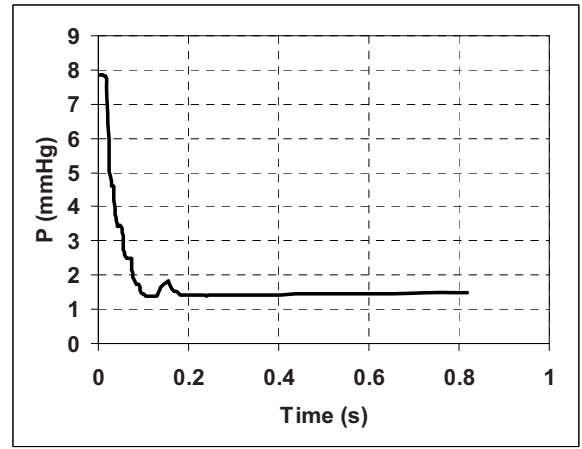

Fig. 9 –Pressure Difference between Brachial and Ascending Aorta

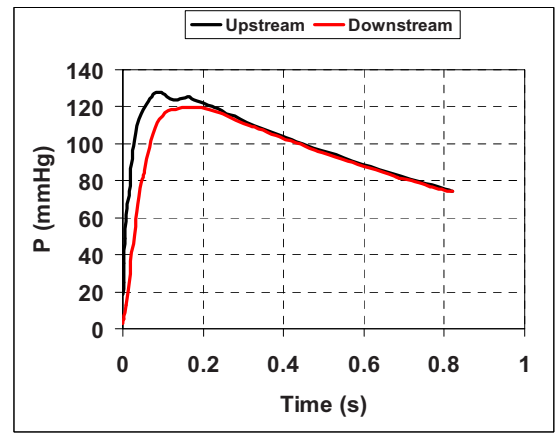

Fig. 10 – Calculated Pressure for Upstream and Downstream (Obstruction) of Ascending Aorta

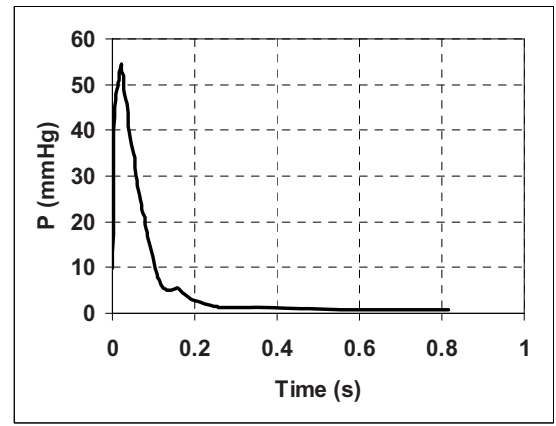

Fig. 11 – Pressure Difference for Obstruction in Ascending Aorta

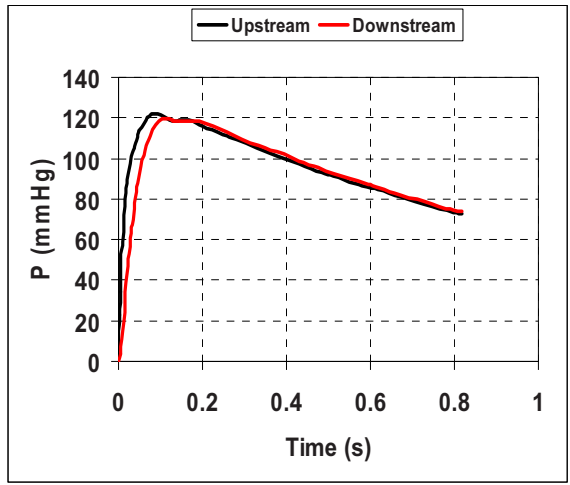

Fig. 12 – Calculated Pressure for Upstream and Downstream (Obstruction) of Right Brachial

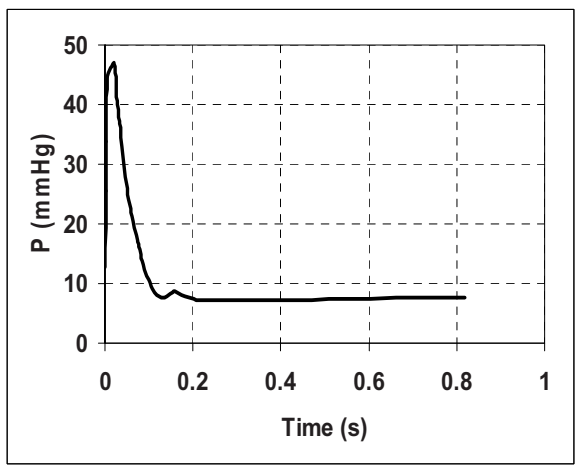

Fig. 13 – Pressure Difference for Obstruction in Brachial

## V. Results

This part contains pressure pulse of ascending aorta and brachial in two cases of normal and abnormal conditions, according to the human cardiovascular simulation.

From Fig7 and Fig8 it is obvious that aortic and brachial pressure graphs are so similar which was explained in upper sections.

Difference between Aortic and brachial pressures is shown in Fig9.

According to above text, we can use verification part to make obstruction results quite citable.

For obstruction in ascending aorta with 27 compartments, results are shown as below. An aortic pressure graph in abnormal condition is shown in Fig10.

Also, Pressure difference between upstream and downstream of obstruction in aorta is shown in Fig11.

Right brachial pressure graphs for brachial in two considered locations are shown in Fig12. Also pressure difference for right brachial because of obstruction is visible in Fig13.

It was said before that beating pressure is approximately in accordance with aortic pressure, but obviously this similarity would be contradicted because of diseases such as obstruction. This contradiction is notable from figures 10 to13.

## VI. Conclusion

At the end it should be noted that using this complex electronic circuit to model human cardiovascular system with its all details, is so useful for studying of blood, different vessels and heart behaviors respect to each other and in different conditions such as health, diseases and abnormities. These abnormities may be obstructions, heart problems, vessels diseases or lots of different other things.

Finally it should be said that our circuit has this tendency to be more accurate and useful by adding more compartments and details to it.

## VII. Refrences